\documentclass[12pt]{article} 
\usepackage{epsf}

 \newcommand{\cP}{{\cal P}}
 
\newcommand{\Z}{{Z \!\!\!  Z}} \newcommand{\eq}[1]{(\ref{#1})}
 
 \newcommand{\beq}{\begin{equation}}
\newcommand{\eeq}{\end{equation}}
\newcommand{\beqn}{\begin{eqnarray}}
\newcommand{\eeqn}{\end{eqnarray}}

\newcommand{\abstracts}[1]{{
\centering{\begin{minipage}{12.2truecm}
\normalsize\baselineskip=15pt \centerline{\footnotesize
ABSTRACT}\vspace*{0.3cm} \parindent=20pt #1 \end{minipage}}\par}}

\begin{document} \begin{flushright} {\large ITEP-TH-60/98}\\
\end{flushright} \vspace{1.5cm}

\begin{center}

{\baselineskip=24pt {\Large \bf Abelian Monopoles and Action
Density}\\ {\Large \bf in $\mathbf{SU(2)}$ Gluodynamics on the
Lattice}\\} {\baselineskip=16pt \vspace{1cm}

{\large B.L.G.~Bakker$^a$, M.N.~Chernodub$^b$,
M.I.~Polikarpov$^b$\\ and A.I.~Veselov$^b$ }\\

\vspace{.5cm} { \it

$^a$ Department of Physics and Astronomy, Vrije Universiteit,\\ De
Boelelaan 1081, NL-1081 HV Amsterdam, The Netherlands

\vspace{0.3cm}

$^b$ ITEP, B.Cheremushkinskaya 25, Moscow, 117259, Russia

} } \end{center}

\vspace{1cm}

\abstracts{We show that the extended Abelian magnetic monopoles in
the Maximal Abelian projection of lattice $SU(2)$ gluodynamics are
locally correlated with the magnetic and the electric parts of the
$SU(2)$ action density.  These correlations are observed in the
confined and in the deconfined phases.}

\vspace{1cm}

\leftline{Pacs:  11.15.H,12.10,12.15,14.80.H}

\newpage

\section{Introduction}

The monopole confinement mechanism~\cite{MatH76} in lattice
gluodynamics seems to be confirmed by many numerical
calculations~\cite{Reviews}.  Monopoles in the Maximal Abelian
(MaA) projection~\cite{MatH76,KrScWi87} are condensed in the
confinement phase of gluodynamics~\cite{MonopoleCond}, their
currents satisfy the classical equations of motion for the (dual)
Abelian Higgs model~\cite{SiBrHa93} and the $SU(2)$ string tension
is reproduced by the monopole currents~\cite{MonopoleDominance}.
The confining string connecting the static qaurk-antiquark pair is
clearly seen~\cite{fluxtube1}.  The next problem to solve is to build the
qualitative and quantitative model for this flux tube, or more
generally the effective infrared Lagrangian for gluodynamics.  The
first steps in this direction are done
already~\cite{fluxtube2,efflagr}. In breef the main results of the
numerical study of the confinement problem are:  the vacuum of
gluodynamics behaves as the dual superconductor, the abelian
monopoles playing the role of the Cooper pairs and the confining
string is an analogue of the Abrikosov-Nielsen-Olesen string.

On the other hand in the continuum theory the Abelian monopoles
arise as singularities in the gauge transformations~\cite{tHo81}.
The definition of the Abelian monopoles is
projection--de\-pen\-dent, monopoles defined in different
projections are different in general\footnote{However, there exists
a gauge invariant definition of the monopole current in any chosen
Abelian projection, see Refs.~\cite{Kyoto96}.}.  Therefore it is
not clear whether these monopoles are "physical" objects.  The
first argument in favour of the physical nature of the Abelian
monopoles was given in Ref.~\cite{ShSu95}:  it was found that the
total action of $SU(2)$ fields is correlated with the total length
of the monopole currents, so there exists a global correlation.
Recently it was shown that the Abelian monopoles in the MaA
projection are locally correlated with the non-Abelian action
density~\cite{BaChPo97}.
Really it means that monopoles are the physical objects (not the
artifacts of the singular gauge transformation), since by
definition we call the object physical if it carries the action.
In Ref.~\cite{Greensite} the correlation
of monopoles, $\Z_2$ strings and the action density was discussed.
The investigation of the correlations of monopoles, the topological
density and the action density was performed in
Refs.~\cite{Markum,MullerPreussker}.

Thus monopoles are important dynamical variables for the
confinement problem and the detailed study of their anatomy is
interesting.  At present we have no idea what is the general class
of the gauge fields which generate the monopole currents in the MaA
projection\footnote{It is known~\cite{InstMon} that instantons
induce Abelian monopole currents in the Abelian gauge but it seems
that they are not the only sources of Abelian monopoles.}.  But
since the elementary monopoles carry nonabelian magnetic action
\cite{BaChPo97} they are related with some nonabelian objects.  The
numerical study of the effective infrared Lagrangian of lattice
gluodynamics shows \cite{efflagr} that to approach the continuum
limit we have to consider also the extended (blocked) monopoles
\cite{IvPoPo}. In the present publication we continue the study of
correlations of the monopole currents and the action density
started in ref.~\cite{BaChPo97}.  We investigate the extended
monopole currents, and also study the correlations of the electric
part of the action with the monopole currents.

The couplings of the monopole action \cite{efflagr} obeys scaling,
it means that these couplings do not depend separately on the
monopole size in the lattice units and bare coupling, but only on
the physical size of the monopole.  This fact in turn means that
the couplings lie on the renormalised trajectory, and we know the
values of the coupling and the size of the monopoles in the
continuum limit ($a \to 0$).  The calculations presented in the
present paper are done just for that sizes of monopoles and for
that values of the bare coupling which correspond to the initial
part ($2.2 < \beta < 2.5$) of the renormalised trajectory of refs.
\cite{efflagr}.  In that sense our results correspond to the
continuum limit.

There are two different types of extended monopoles (type-I and
type-II monopoles)~\cite{IvPoPo}.  As we already discussed the
type-II extended monopoles are important dynamical variables in
lattice gluodynamics~\cite{efflagr,ShSu95}.  The type-I extended
monopoles play a non-trivial role for the dynamics of the phase
transitions in electroweak theory~\cite{ChGuIlSch} and in the
$U(1)$ Abelian Higgs model~\cite{Kajantie}.

The paper is organised as follows.  In Section~2 we introduce two
quantities $\eta^{\mathrm{E}}$ and $\eta^{\mathrm{M}}$, which
define the correlation of the magnetic and electric parts of the
$SU(2)$ action with the (extended) monopole charge.  In Section 3
we describe the results of numerical calculations.  We discuss the
results in Section~4.

\section{Correlations of Monopoles with Action Densities}

If the Abelian monopole carries the non-Abelian action, then the
action density near the monopole current should be larger than the
action density far from the monopole trajectory.  One of the
quantities which can show this effect is the relative excess of the
mean action density in the region near the monopole
current~\cite{BaChPo97}.  The total action can be divided into
electric and magnetic parts.  The relative excess of the magnetic
(electric) action density is defined as:  \beqn
\eta^{\mathrm{M(E)}} = \frac{S^{\mathrm{M(E)}}_m - S}{S}\,.
\label{eta} \eeqn Here $S = \langle S_P \rangle \equiv \langle
\left( 1 - \frac 12 Tr \, U_{P}\right) \rangle $ is the expectation
value of the lattice plaquette action.  The quantity
$S^{\mathrm{M}}_m$ is the action averaged over the plaquettes
closest to the monopole current $j_\nu(x)$.  The definition of
$S^{\mathrm{M}}_m$ is:  \beqn S^{\mathrm{M}}_m= \langle \frac 16
\sum_{P \in \partial C_\nu (x)} S_P \rangle \,, \label{Smlat} \eeqn
where the summation is over the plaquettes $P$ which are the faces
of the cubes $C_\nu (x)$; a cube $C_\nu (x)$ is dual to the
monopole current $j_\nu(x)$.  For the static Abelian monopole $j_0
(x) \neq 0,\, j_i (x) =0$ ($i=1,2,3$), and the boundaries of the
cubes dual to the monopole current are formed by the space--like
plaquettes $P_{i,j}$, $i,j=1,2,3$.  Therefore only the magnetic
part of the $SU(2)$ action density, $1 \slash 2 \, {\mathrm{Tr}
F^2_{ij}}$, contributes to $S^{\mathrm{M}}_m$.

The quantity $S^{\mathrm{E}}$in eq.~\eq{eta} is:  \beqn
S^{\mathrm{E}}_m= \langle \frac{1}{24} \sum_{P \in \cP(C_\nu (x))}
S_P\rangle \,, \label{Selat} \eeqn where $\cP(C_\nu(x))$ is the set
of all plaquettes $P$ which satisfy the following two conditions:
all plaquettes $P$ {\it (i)}~have one, and only one, common link
$l_\mu$ with the cube $C_\nu (x)$; {\it (ii)} they are lying in the
planes, defined by the vectors $\hat \mu$ and $\hat \nu$.  There
are 24 such plaquettes corresponding to a cube $C_\nu(x)$.  For the
static monopole current these plaquettes lie in the planes $(0,i)$,
$i=1,2,3$; therefore only the electric part of $SU(2)$ action
density, $1 \slash 2 \, {\mathrm{Tr} F^2_{0i}}$, contributes to the
quantity $S^{\mathrm{E}}_m$.

Thus, our definition of electric, $S^{\mathrm{E}}$, and magnetic,
$S^{\mathrm{M}}$, parts corresponds to the electric and magnetic
parts of the action density only for a static monopole.  For
non-static monopoles it is convenient to keep these notations.

In the naive continuum limit the expressions \eq{Smlat} and
\eq{Selat} and the plaquette action $S$ have the following form:
\beqn S^{\mathrm{M}}_m & = & \frac{1}{24} \langle {\rm Tr}{
\Bigl(n_\mu (x)\,{\tilde F}_{\mu\nu}(x)\Bigr)}^2 \rangle
\label{Smc}\,,\\ S^{\mathrm{E}}_m & = & \frac{1}{6} \langle {\rm
Tr}{\Bigl(n_\mu(x)\, F_{\mu\nu}(x)\Bigr)}^2 \rangle
\label{Sec}\,,\\ S & = & \frac{1}{24} \langle {\rm Tr} F^2_{\mu\nu}
\rangle \,, \label{act} \eeqn where ${\tilde F}_{\mu\nu} = 1 \slash
2 \varepsilon_{\mu\nu\alpha\beta} F_{\alpha\beta}$, and $n_\mu (x)$
is the unit vector in the direction of the current:  $n_\mu
(x)=j_\mu (x)/| j_\mu (x) |$ if $j_\mu (x) \neq 0$, and $n_\mu (x)=
0$ if $j_\mu (x) = 0$.  For a static monopole eq.~\eq{Smc} and
eq.~\eq{Sec} give the normalised average of the chromomagnetic,
$1\slash 3 {(B^a_i)}^2$, and chromoelectric, $1\slash 3
{(E^a_i)}^2$, action density at the point where the monopole is
located.  Eq.~\eq{act} gives the normalised total action density:
$ <S> = 1\slash 6\, <{(B^a_i)}^2+ {(E^a_i)}^2>$.

\section{Numerical Results}

Below we present the quantities $\eta^{\mathrm{M}}$ and
$\eta^{\mathrm{E}}$ calculated on symmetric, $24^4$, and
asymmetric, $24^3 \cdot 4$ lattices in standard $SU(2)$ lattice
gluodynamics \footnote{To check the finite volume corrections we
also performed calculations on the smaller lattices:  $16^4$,
$20^4$, $16^3 \cdot 4$ and $16^3 \cdot 4$.  It occurs that the
results obtained on these small lattices coincide within the
statistical errors with the results obtained on $24^4$ and $24^3
\cdot 4$ lattices}.  In all these cases, we find in the MaA
projection~\cite{KrScWi87} that the quantities
$\eta^{\mathrm{M,E}}$ are different from zero for all values of
$\beta$.  We also considered the $F_{12}$ (diagonalization of the
$F_{12}$ lattice field strength tensor), Polyakov (diagonalization
of the Polyakov line) and ``random'' Abelian gauges.  The
``random'' Abelian gauge means no gauge fixing at all:  we take a
field configuration, apply a random gauge transformation and then
treat the phases of the diagonal elements of the $SU(2)$ gauge
field as the Abelian gauge field.

To fix the MaA projection we use the overrelaxation algorithm of
Ref.~\cite{MaOg90}.  The number of gauge fixing iterations is
determined by the following criterion~\cite{Pou97}:  the iterations
are stopped when the matrix of the gauge transformation $\Omega
(x)$ becomes close to the identity matrix:  ${\rm max}_x \{ 1-
{\frac 12} Tr \,\Omega (x) \} \le 10^{-6}$.  We also check that a
more accurate gauge fixing does not change our results.  By
performing a sufficient number of iterations between measurements
we have made sure that the configurations on which we performed our
measurements are statistically independent.

\begin{figure}[!htb] 
\begin{center} 
\begin{tabular}{cc}
\hspace{-0.8cm}\epsfxsize=7.2cm\epsffile{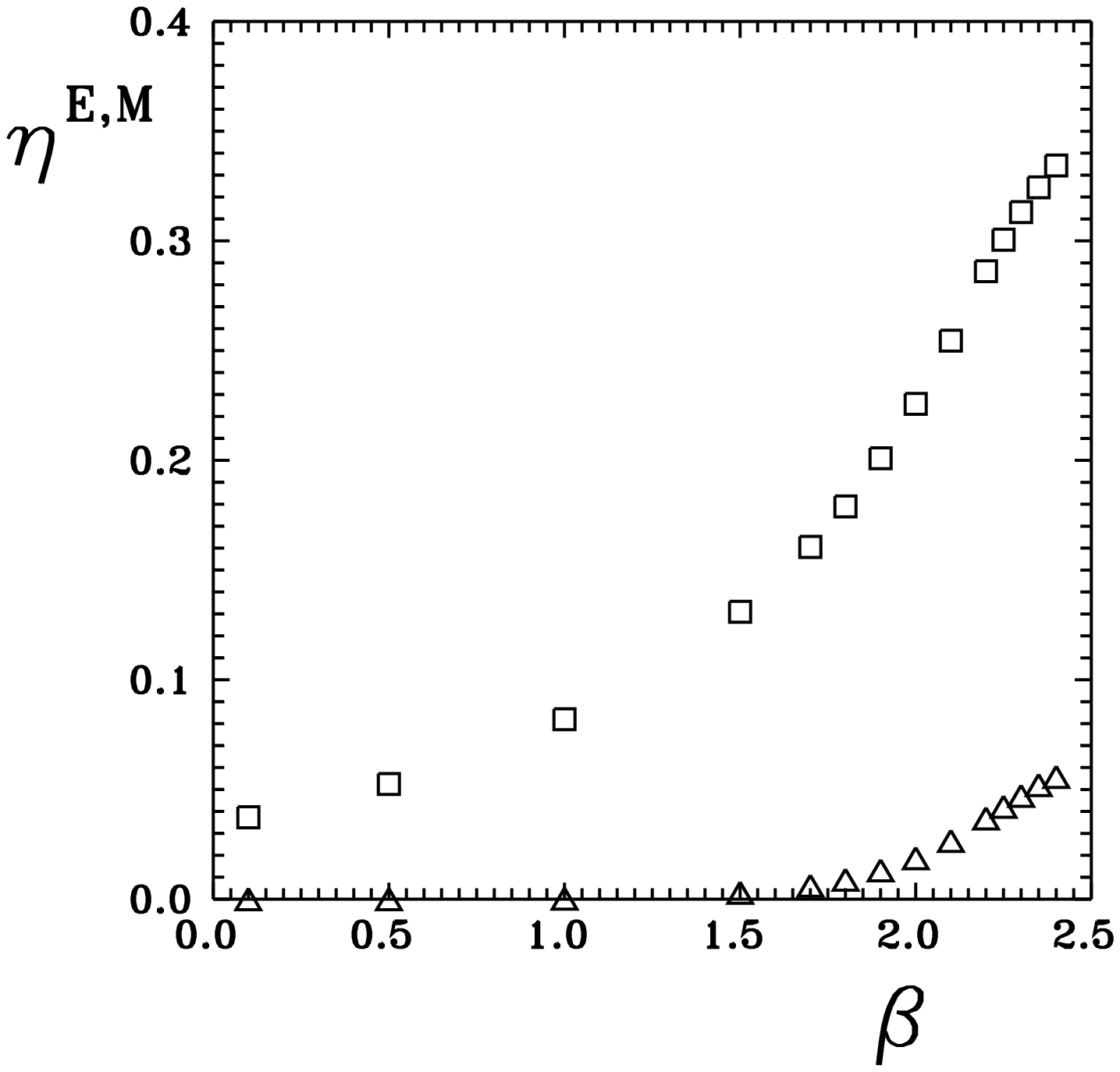} &
\hspace{0.8cm}\epsfxsize=7.2cm\epsffile{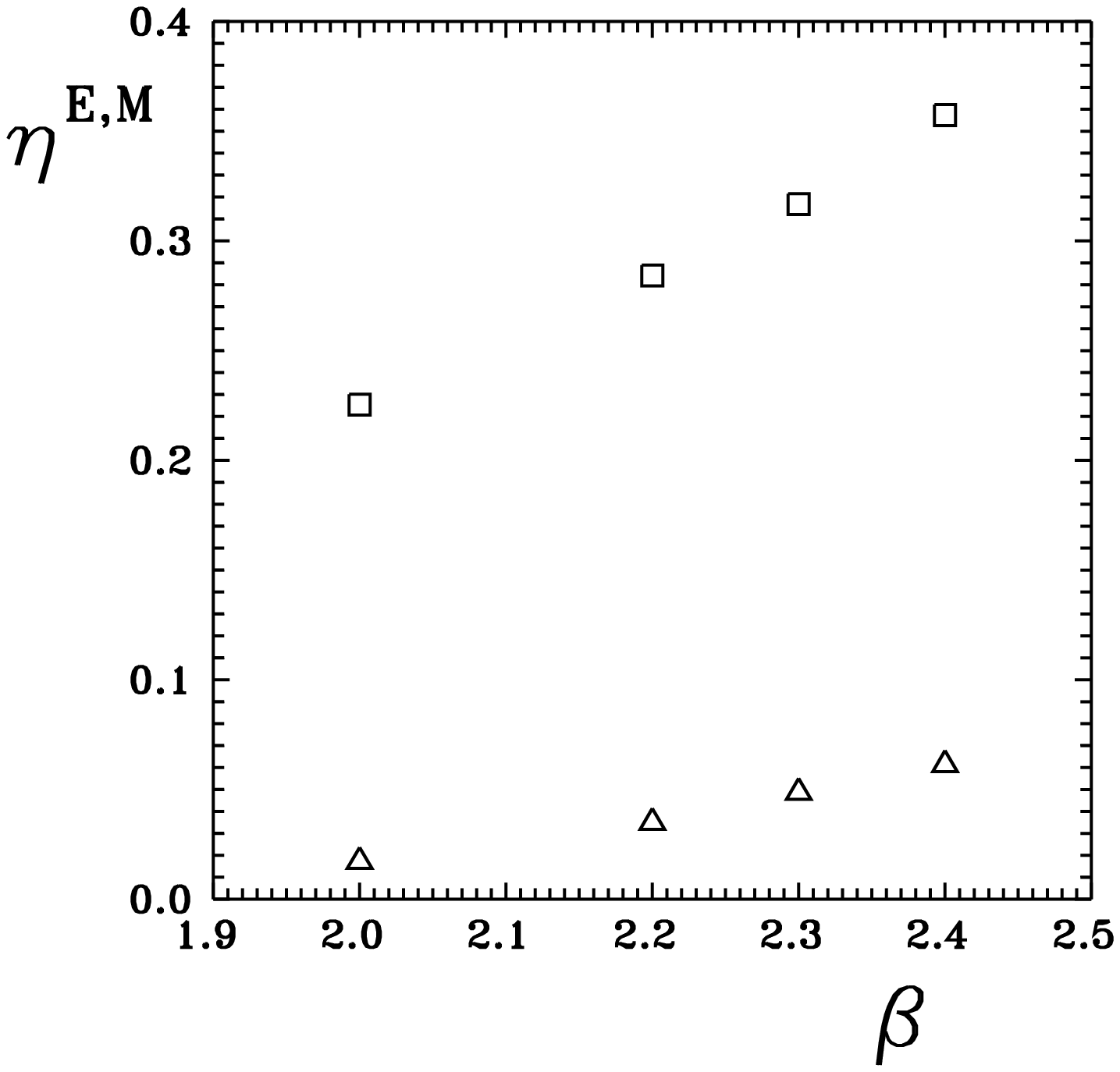} \\ (a) &
\hspace{1.5cm} (b) \vspace{0.5cm}\\ 
\end{tabular} 
\end{center}
\vspace{-1cm} 
\caption{(a) The quantities $\eta^{\mathrm{M}}$
(boxes) and $\eta^{\mathrm{E}}$ (triangles) $vs.$ $\beta$ on the
lattice $24^4$ for the MaA projection.  In all figures the error
bars are much smaller than the sizes of the symbols used; (b) the
same as in (a), but now for the lattice $24^3\cdot 4$.}
\label{one} 
\end{figure}

Figure~\ref{one}(a) shows the quantities $\eta^{\mathrm{M,E}}$ in
the MaA projection for the lattice $24^4$.  The quantity
$\eta^{\mathrm{M}}$ is $4-6$ times larger than the quantity
$\eta^{\mathrm{E}}$ for all considered values of $\beta = 4 \slash
g^2$.  Thus the excess of the chromomagnetic action near the
monopole position is larger than the excess of the chromoelectric
action.

The correlations increase with increasing $\beta$.  For small
$\beta$ monopoles are present almost everywhere, so the action
averaged over the cubes containing monopoles differs very little
from the action averaged over all cubes.  The density of monopoles
decreases with increasing $\beta$, thus the increase of the
correlator as $\beta \to \infty$ means that at large $\beta$ the
Abelian monopoles disappear mainly in the regions with a small
$SU(2)$ action density.

Note, that the monopole current $j_\mu$ is derived from the
plaquettes $\partial C_\mu$ which contribute to $S^{\mathrm{M}}$,
thus the fact that $\eta^{\mathrm{M}} \neq 0$ is rather natural.
The plaquettes which contribute to $S^{\mathrm{E}}$ are not
directly related to the monopole current and the fact that
$\eta^{\mathrm{E}} \neq 0$ probably means that there exist some
structures in the vacuum of gluodynamics which generate monopole
currents and carry electric and magnetic action.

Our numerical simulations show that the quantities
$\eta^{\mathrm{E,M}}$ for the $F_{12}$, Polyakov and random Abelian
gauges coincide with each other within the numerical errors.  For
all studied values of the coupling constant $\beta$ the values of
the quantities $\eta^{\mathrm{M,E}}$ calculated in these gauges are
more than $10$ times smaller than those for the MaA gauge.  This
fact probably indicates that the Abelian monopoles in the $F_{12}$
and Polyakov Abelian projections carries much less information
about the properties of the non-Abelian vacuum than the Abelian
monopole in the MaA projection\footnote{In
Ref.~\cite{MullerPreussker} the correlator $\eta^{\mathrm{M}}$ is
studied under the smoothing procedure.  It was found that for
elementary monopoles in different gauges this correlator is of the
same order.  The smoothing procedure removes short-range
fluctuations, therefore the result of Ref.~\cite{MullerPreussker}
probably indicates that the small correlation of the monopoles with
the action density in, say, the Polyakov gauge, is due to
ultraviolet vacuum fluctuations.}.

The finite-temperature analysis of the correlators
$\eta^{\mathrm{M,E}}$ is performed on an asymmetric lattice.  We
found that at finite temperature the correlators in the MaA
projection turned out to be much larger than the correlations in
the $F_{12}$, Polyakov and random gauges.  We show the quantities
$\eta^{\mathrm{M}}$ and $\eta^{\mathrm{E}}$ in the MaA gauge in
Figure~\ref{one}(b).  These calculations are performed on a $24^3
\cdot 4$ lattice.  It is seen that the confinement--deconfinement
phase transition (which occurs at $\beta = \beta_c = 2.3$) has no
observable influence on the behaviour of the correlators
$\eta^{\mathrm{M,E}}$.

The correlation between electric and magnetic action in the
vicinity of Abelian monopoles is small.  We measured the
correlation of the product of the electric and the magnetic action
with the monopole currents.  We find that the correlator
\begin{equation} \eta^{\mathrm{EM}} = \frac{\langle
S^{\mathrm{E}}(x) S^{\mathrm{M}}(x) \rangle}{ \langle
S^{\mathrm{E}}(x) \rangle \langle S^{\mathrm{M}}(x) \rangle } - 1
\label{eqB.1} \end{equation} vanishes within the statistical error
for the studied region of the bare coupling $\beta$ on the lattice
$24^4$.  This occurs not only when the averages are taken over the
full lattice, but also if only the cubes associated with monopoles
are included in the average.  This result is independent of the
gauge fixing condition.  Therefore the magnetic and electric
fluctuations around the Abelian monopole in the MaA gauge are
independent.

\begin{figure}[!htb] 
\begin{center}
\begin{tabular}{cc}
\hspace{-0.8cm}\epsfxsize=7.2cm\epsffile{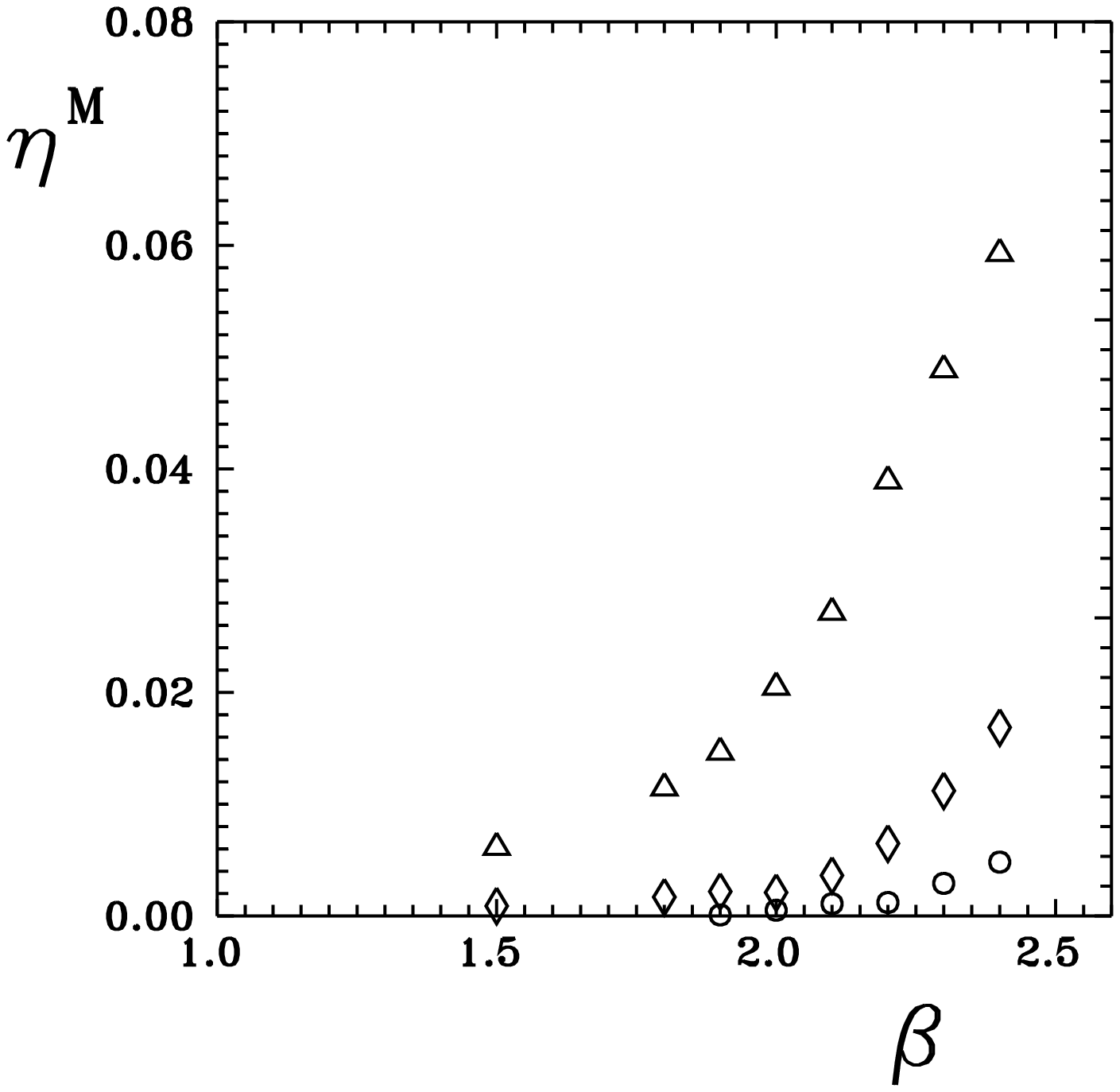} &
\hspace{0.8cm}\epsfxsize=7.2cm\epsffile{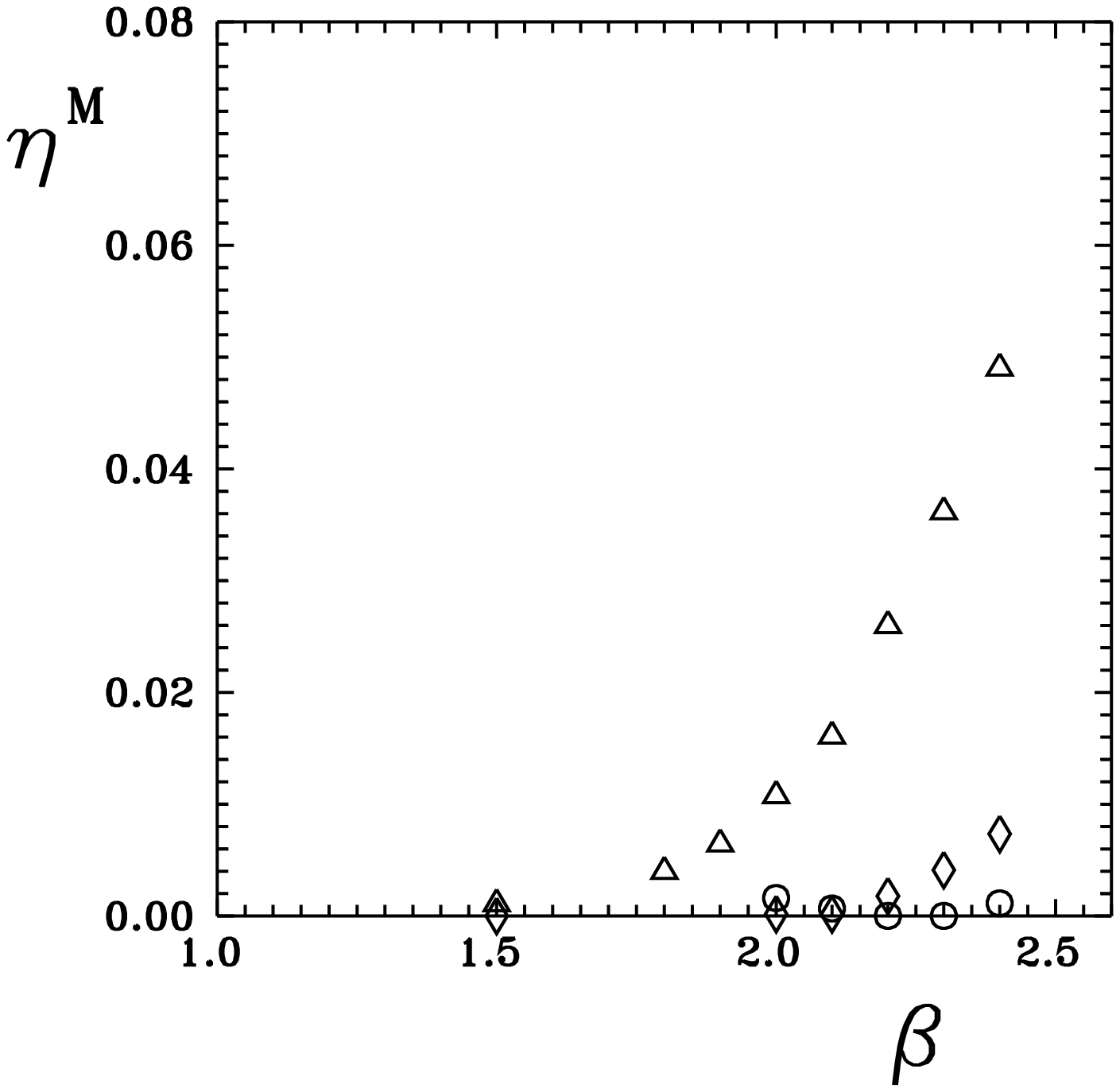} \\ (a) &
\hspace{1.5cm} (b) \vspace{0.5cm}\\ \end{tabular} \end{center}
\vspace{-1cm} \caption{(a) The correlator $\eta^{\mathrm{M}}$ in
the MaA gauge for a $24^4$ lattice $vs.$ $\beta$ for type-II
extended monopoles of sizes 2 (triangles), 3 (diamonds) and 4
(circles); (b) The same as in (a), but now for type-I monopoles.}
\label{two}

\end{figure}

\begin{figure}[!htb]
\centerline{\epsfxsize=0.90\textwidth\epsfbox{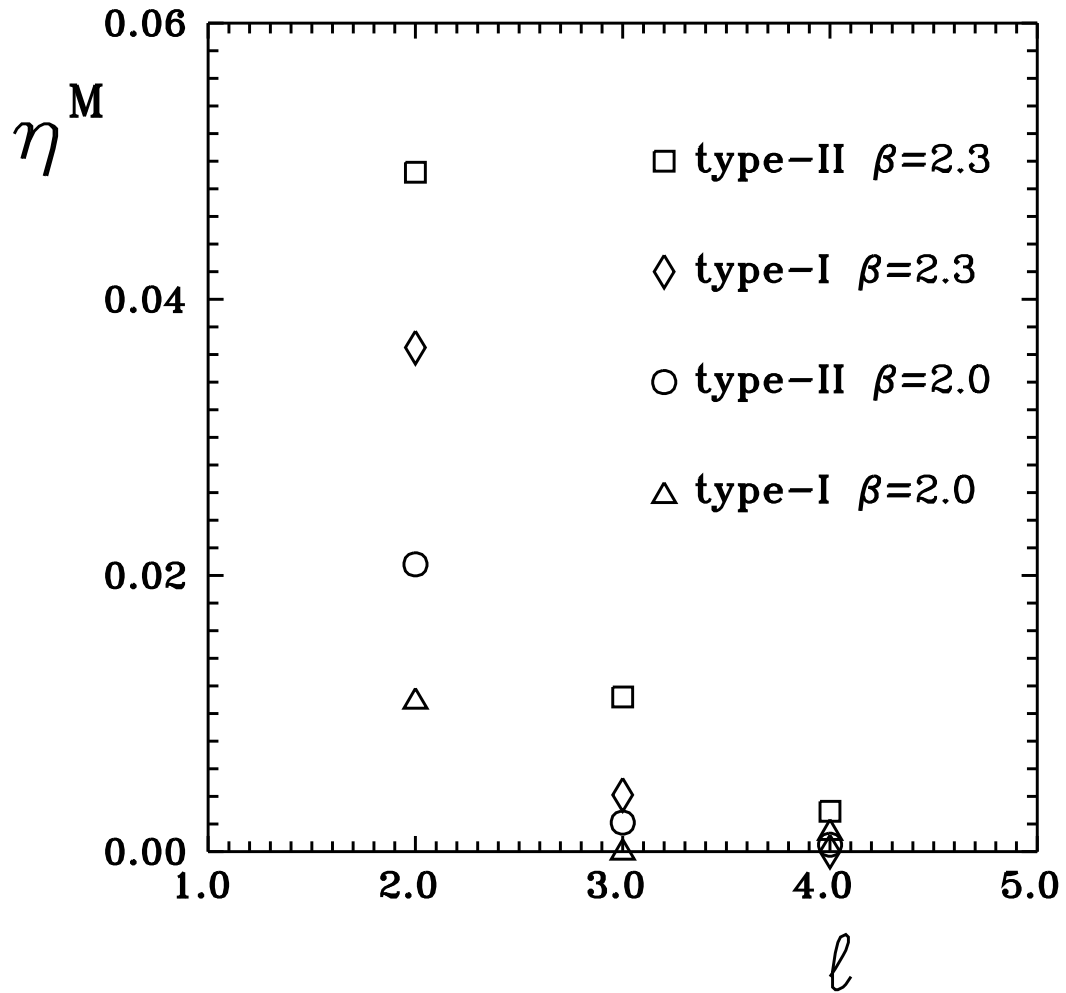}}
\caption{The correlator $\eta^{\mathrm{M}}$ plotted as a function
of the linear dimension $\ell$ of the extended monopoles, the
lattice size is $24^4$.}  \label{three} \end{figure}

We also study the correlations of extended monopoles~\cite{IvPoPo}
with the electric and magnetic action.  There are two types of
extended monopoles~\cite{IvPoPo}:  type I corresponds to the
plaquettes of size $\ell \times \ell$; type II uses all $1\times 1$
plaquettes that tile the faces of an $\ell^3$ sized cube associated
with a monopole current.  We measured the correlations of the
magnetic and the electric action densities with the extended
monopoles of sizes $\ell = 2$, 3 and 4.  It turns out that for the
whole range of the bare coupling $\beta$ studied, the quantity
$\eta^{\mathrm{M}}$ for type-II monopoles is larger than that for
the type-I monopoles.  In Figures~\ref{two}(a,b) we show the
dependence of the quantity $\eta^{\mathrm{M}}$ on $\beta$ for
type-II monopoles and type-I monopoles.  In order to show a
similarity between different types of monopoles we plot the
quantity $\eta^{\mathrm{M}}$ for the type-I and type-II monopoles
$vs.$ linear size of the extended monopoles $\ell$
(Figure~\ref{three}).  The figure clearly shows that the larger the
size of the monopole the smaller the correlation
$\eta^{\mathrm{M}}$ is.  This fact is not unexpected since with
increasing monopole size the part of the lattice which belongs to a
monopole gets larger and therefore the averaged action associated
with the monopole gets closer to the total averaged action.  If the
correlations $\eta^{\mathrm{M}}$ and $\eta^{\mathrm{E}}$ are
physical quantities then they should depend on the physical
monopole size, $b = \ell a$, where $a$ is the physical lattice
spacing.  We are planning to study this dependence in our next
publication.

\section{Discussion and Conclusions}

We discussed the local correlations of the electric and magnetic
parts of the $SU(2)$ action with Abelian monopoles in various
Abelian projections.  We have shown that monopoles in the Maximal
Abelian projection are correlated with the electric and magnetic
parts of the action density at zero and at finite temperature.  The
same result is obtained also for type-I and type-II extended
monopoles.  The correlators $\eta^{\mathrm{M,E}}$ for the type-II
monopoles are always larger than the correlators for the type-I
monopoles.  Thus, for the description of the vacuum of $SU(2)$
gluodynamics the type-II monopoles are more suitable variables than
the type-I monopoles, in agreement with
Refs.~\cite{ShSu95,efflagr}.

The correlation of the monopoles with the electric part of the
action density is smaller than the correlation with the magnetic
part of the action density.  The correlations of the Abelian
monopole with both parts of the $SU(2)$ action density in the
Polyakov, $F_{12}$ and random gauges are of the same order; all of
them are much smaller than the correlations in the MaA gauge.

We note here that the existence of the correlation of the electric
and the magnetic action densities with the Abelian monopoles can be
understood from the fact that the Abelian monopoles are correlated
with the topological charge density~\cite{Markum,jejm1,jejm2}.
Indeed this correlation means that the monopole currents are
accompanied by a non-zero density of the topological charge.  This
charge is non-zero if and only if both the electric and magnetic
action densities are non-zero.

We conclude that the Abelian monopoles in the Maximal Abelian
projection are physical objects which carry both magnetic and
electric parts of the $SU(2)$ action density.

\section*{Acknowledgements}

The authors are grateful to T.~Suzuki and Yu.A.~Simonov for useful
discussions.  M.N.Ch.  and A.I.V.  feel much obliged for the kind
reception given to them by the staff of the Department of Physics
and Astronomy of the Free University at Amsterdam.  This work was
partially supported by the grants INTAS-96-370, INTAS-RFBR-95-0681,
RFBR-96-02-17230a, RFBR-97-02-17491a and RFBR-96-15-96740. The work of
M.N.Ch. was supported by the  INTAS Grant 96-0457 within the research
program of the International Center for Fundamental Physics in Moscow.

\newpage

\section*{Figures}


\end{document}